\documentclass[fleqn,10pt]{wlscirep}
\usepackage[utf8]{inputenc}
\usepackage[T1]{fontenc}

\usepackage{amsmath,algorithm, algpseudocode,multirow, amssymb,amsfonts,graphics,graphicx,dcolumn,bm,enumerate}

\title{Effect of lockdown interventions to control the  COVID-19 epidemic in India}

\author[1]{Ankit Sharma}
\author[1]{Shreyash Arya}
\author[1]{Shashee Kumari}
\author[1]{Arnab Chatterjee}
\affil[1]{TCS Research, New Delhi, India}

\begin{abstract}
The pandemic caused by the novel Coronavirus SARS-CoV2 has been responsible for life threatening health complications, and extreme pressure on healthcare systems.
While preventive and definite curative medical interventions are yet to arrive, Non-Pharmaceutical Interventions (NPIs) like physical isolation, quarantine and drastic social measures imposed by governing agencies are effective in arresting the spread of infections in a population. In densely populated countries like India, lockdown interventions are partially effective due to social and administrative complexities. Using detailed demographic data, we present an agent based model to imitate the behavior of the population and its mobility features, even under intervention.
We demonstrate the effectiveness of contact tracing policies and how our model efficiently relates to empirical findings on testing efficiency. We also present various lockdown intervention strategies for mitigation -- using the bare number of infections, the effective reproduction rate, as well as using reinforcement learning. Our analysis can help assess the socio-economic consequences of such interventions, and provide useful ideas and insights to policy makers for better decision making.
\end{abstract}
\begin{document}

\flushbottom
\maketitle
%
%
\thispagestyle{empty}

\section{\label{sec:intro}Introduction}

The rapidly spreading infectious disease COVID-19 is caused by the novel coronavirus SARS-CoV-2, has become a global pandemic~\cite{WHOdg}. Spreading fast, with short doubling times there are usually long delays in showing effects from interventions~\cite{pellis2020challenges}. With increasing number of complex, life threatening infections~\cite{guan2020clinical}, it has overwhelmed healthcare systems around the world and caused thousands of deaths worldwide. A majority of infected individuals need medical treatment, some of them critical attention. The current case fatality ratio (CFR) stands around 7\%, although countries across the world show variations between 1\% and 15\%~\cite{JHUmortality}.
While the World Health Organization (WHO) currently maintains that there are no reports of reinfection yet, they were cautious in stating that there is \textit{``currently no evidence that people who have recovered from COVID-19 and have antibodies are protected from a second infection''}~\cite{WHOreinf}.

The propagation of a transmitting infection is usually well understood by the basic reproduction number $\rho_0$ which is the average number of infections caused by a single individual in a fully susceptible population, or in other words, a population without any immunity. When immunity develops in a population, or otherwise, when other mitigating factors are in play, the effective reproduction number $\rho_t$ (reproduction number when both susceptible and non-susceptible are present) decreases to below unity, thereby rapidly decreasing the proportion of infected population. Suppressing infections is an alternative way of stopping the spreading, since the population still remains susceptible. True immunity is only achieved using pharmaceutical intervention like vaccination.
In a well mixed population, the herd immunity is either achieved by naturally recovering from infections or by vaccinations, when $1-\frac{1}{\rho_0}$ fraction of population becomes immune.

In absence of vaccines and consensus on effective medication for treatment, drastic \textit{Non-Pharmaceutical Interventions} (NPIs) are the only possible ways to control the contact mixing of the population that is the basis of the spreading of any contact mediated infectious disease. The effect of such interventions only delay the spread of the infections in the susceptible population, thereby reducing the pressure on the healthcare system, and buying time for viable, effective pharmaceutical answers.

One of the basic and widely used intervention is to impose strict restriction on population contact, mixing and movement, usually regulated by governing authorities, and are commonly termed as \textit{lockdowns}. When the duration of lockdowns keeps increasing, the public costs can be immense, both in terms of its economic and financial fall-outs, as well as from the social and psychological perspective. Recent proposals of isolation and other  restrictions~\cite{ferguson2020report} have barely been able to \emph{flatten the curve} and keep the critical cases below the healthcare capacity. Flattening the curve along with managing social and economic costs is not possible unless NPIs are introduced and relaxed to re-start the social and economic activities in a civil society.

One of the effective ways of monitoring and controlling the spread of infections is through the efficient use of instantaneous contact tracing data~\cite{ferretti2020quantifying} acquired through smartphone based application (\textit{app}), and administer regulated interventions accordingly. The \textit{app} keeps track of proximity contacts and notifies individuals at risk, who can isolate, quarantine or get tested and treated. Such targeted recommendations can serve as a better alternative for infection control, compared to mass quarantines, which can have heavy socio-economic consequences. However, this can only be possible if the application can be used by a large fraction of the population. A recent study~\cite{bulchandani2020digital} argued that while almost 87\% of Indian population has access to a mobile phone~\cite{TRAI}, the number of smartphone users is still well below 40\%. If there is a mechanism to augment the location data from non-smartphone users using basic feature phones, the Indian population can \textit{achieve digital herd immunity}, thereby facilitating the government authorities to track individuals, trace their contacts and impose targeted interventions like isolation, testing and quarantine. However, such a theoretical possibility is yet to mature to a practical reality.

Modeling epidemics on graphs~\cite{pastor2015epidemic} has been an extremely important and fertile area of research. The natural reason is to understand the mechanisms of disease propagation in closely connected and mixing population in human societies, using a variety of frameworks from paradigmatic toy models to detailed, data driven, agent based models. The huge amount of data gathered through a multitude of sources to create a multi-scale data of demographics, mobility and other essential components have opened up the possibility to perform large scale data-driven 
simulations. This helps to answer important, specific and detailed questions related to epidemics, and particularly for prediction and forecasting purposes. 

In this paper, we introduce an agent based model of the Indian population, with demographic and mobility features modeled using publicly available data sources. The COVID-19 epidemic is studied using a microscopic model that takes into account the available data on transmission, infections and mortality. Under this detailed framework, we study the consequences of a few lockdown strategies that can be effective in containing the spread of infections as well as possibly minimize the socio-economic footprints.
In the following section, we describe our model framework in detail by introducing the structured population and then the epidemic process. Next, we discuss our results and conclude with discussions.

\section{\label{sec:model}Model}
We separate our model into two distinct parts. First of all, we model the population and its contacts as a hierarchically structured graph, in order to facilitate the effect of lockdown intervention at different spatial scales by rearranging and removing links systematically. The  movement related behavior of individuals depend on several demographic features. The second part of the model deals with the epidemic dynamics and its details.

\subsection{\label{subsec:pop}Modeling the population}
We build a hierarchically modular network to simulate a structured population (See, e.g. Ref.~\cite{davis2020phase}) where the basic units are individuals. While the individual agents are the nodes, the links represent the close physical proximity or contact and associations over a short period of time, which can be dynamic in nature. We will interchangeably use the term \emph{agents} to refer to individuals in the rest of the paper, since they will eventually have several static and dynamic features depending on spatial and temporal properties.
This can be imagined to be the aggregated network of all contact graphs of agents over a short period of time.
The modular graph is conceived and constructed using the following steps:
\begin{figure}[h]
    \centering
    \includegraphics[width=0.7\linewidth]{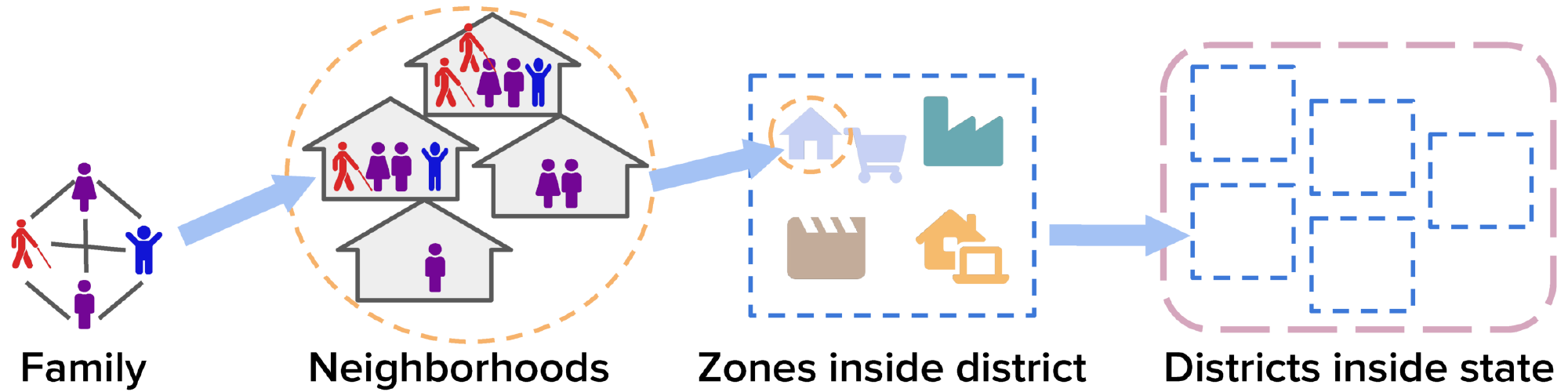}
    \label{fig:hierarchy}
    \caption{The hierarchical structure of the population: the basic units are individuals, forming families residing in residential zones which also contain retail. The contact networks between families are governed by Barab\'{a}si-Albert type networks. There are also zones for education and entertainment, as well as employment (public or private) opportunities. Combinations of these zones create district. Travel between various areas create contact networks between individuals. Several such districts constitute a state. Contact networks between districts are created using gravity law.}
\end{figure}

\begin{itemize}
    \item 
    Individuals (agents) form the basic units of a society. However, at the basic level, several of them constitute a family. It is fair to assume that individuals in a family form a complete graph or a clique, because they come in close contact in their residential space. 
    
    \item 
    In a residential neighborhood, several families live in close proximity. It is fair to assume that number of contacts with others across different families have a broad distribution. In fact, there are strong empirical evidences that the degree distribution of physical social networks have power law tails (see, e.g.~\cite{eubank2004modelling}).
    We construct a neighborhood using different families as units. We use the Barab\'{a}si-Albert algorithm for preferential attachment~\cite{barabasi1999emergence} to connect a new family of size $n_k$ to an existing family $i$, with probability $n_i/\sum_j n_j$, where $n_j$ are the sizes of existing families. Once a target family is selected, we create a single link between any two members of the two families to keep the connectivity sparse. The family size distribution is sampled from the Census data~\cite{census2011}.
    See Appendix for details in Algorithm~\ref{alg:CreateNetwork}.
    
    \item 
    Next, we introduce the concept of \emph{zones} as a spatial feature. Each zone is tagged to some special amenities like residential neighborhoods (defined in the previous paragraph), retail outlets, entertainment centers, educational institutions, office spaces, etc., which provide distinguishing characteristic to it. 
    In a simplistic picture, we assume that there are 4 different types of zones in a district: 
    (i) residential and retail (RR), (ii) education and  entertainment (EE), (iii) private employment (EPV), and (iv) public employment (EPB). 
    A district usually comprises of several of the above zones. Zones in each district are assumed to be distributed uniformly for sake of simplicity. For instance, Mumbai Suburban district has a total of 25 zones divided in ratio 8:5:5:7 for RR:EE:EPV:EPB respectively.
    
    \item
    Agents are embellished with various features from the demographic data available from Census of India 2011~\cite{census2011}. We use age, gender (male or female) and employment status. We consider 3 broad age classes -- (i) 0-19 years, (ii) 20-59 years and (iii) 60 years and above. We also have employment status as -- (i) employed or (ii) unemployed. The unemployed do not move to the employment zones. The employed in turn also fall in two categories -- (i) essential and (ii) non-essential workers. The essential workers have the liberty to move around different zones even during the lockdown. For instance, 48.73\% of  Mumbai Suburban's population is employed out of which 1\% of agents are tagged as essential workers randomly (See Appendix Table~\ref{tab:param_notation1} for parameter details,  $F_{essential}$). 
    
    \item 
    We consider 4 time bins -- (i) morning, (ii) afternoon, (iii) evening and  (iv) night, which sets the time granularity of our model. Thus, 4 time steps correspond to one day in our subsequent simulations.
    
    Inspired by models of intra-urban mobility (see, e.g., Ref~\cite{serok2015simulation}), we simulate the movement of agents across different zones as a function of their age and time of the day, using probabilities for transiting from one zone to another. We also construct a similar set of conditional probability tables for the lockdown periods taking ideas from ~\cite{serok2015simulation, Google_mobrep}, when movement is rarely allowed outside residential zones except for obtaining essential commodities. 
    
    At this point it is important to note that the mobility data at the level of individuals is useful, to compute origin-destination matrices, dwell estimates at hotspots, amount of time spent at various locations, and also  contact matrices~\cite{oliver2020mobile}. Compared to other countries like Italy~\cite{pepe2020covid}, the data at this
    level of detail is not available for India.
    However, aggregate mobility trends~\cite{Google_mobrep} give a rough idea of the time aggregated mobility across a variety of locations like retail and recreation, groceries and pharmacy, parks, transit stations, workplaces and residential areas.
    See Appendix Table~\ref{tab:mobility matrix1}, Table~\ref{tab:mobility matrix2}, and Table~\ref{tab:mobility matrix3} for details. 
    
    \item
    Restructuring the network due to mobility: 
    We model the movement of agents across different zones and the resulting contact structure with others, using a process of removing and adding links in the existing graph.
    If an agent does not move out of its own zone, it breaks one of its existing links with probability $P_{re}$ and creates a new link inside the zone with probability $P_{ce}$. In case if it moves out of its own zone to a target zone (can be within or outside its own district), all links from its previous zone are broken with a probability $P_{re}$ and a new link is created in the target zone with probability $P_{ce}$. An agent moves to another district as per the inter-district mobility matrix $M_{DD}$. This dynamics of links effectively captures the properties of temporal contact graphs at the granularity of 4 time steps in a single day. At night, agents return back to their residence locations with a probability $C_H$, restores their family links with probability $P_{ce}$, and their existing links are removed with a probability $P_{fe}$ . As soon as an agent is tested positive, its existing links are broken with probabilities $C_I$ and $C_E$ depending upon the current state of agent i.e., Infected (symptomatic) or Exposed (asymptomatic) respectively (See Sec.~\ref{subsec:epid} for epidemic states). If an agent dies, all its links are broken and the agent is removed from the system. To restrict/limit inter-district mobility, inter-district links are broken with a probability $C_{BS}$ when lockdown is imposed. During lockdown, parameters are changed to bring in the effect of social distancing. Refer to Table~\ref{tab:param_notation3}, Table~\ref{tab:param_notation4}, and  Algorithm~\ref{alg:MobilityModel} in Appendix for details.

    \item 
    Connections across districts:
    Empirical evidences regarding population movement across space is a well studied and debated problem. While earliest research pointed the flux between two population agglomerations to be proportional to the respective population sizes and a inverse dependence on the distance separating them~\cite{tinbergen1962shaping,ravenstein1889laws}, recent studies present strong evidence in favour of a generalized gravity law~\cite{balcan2009multiscale,viboud2006synchrony}, which holds reasonably well depending on the scale under consideration. In absence of empirical data for population flux in India, we assume that the gravity law holds, as in the case of Korean highways~\cite{jung2008gravity}. 
    
    We assume that the number of links across districts is directly proportional to the population flux between two districts of population sizes $P_i$ and $P_j$, separated by distance $d_{ij}$, and is given by 
    \begin{equation}
        F_{ij} = C \frac{P_i P_j}{d_{ij}^2}.
    \end{equation}    
    The normalization factor $C$ is set to unity. In each district, nodes are chosen randomly to connect with randomly chosen nodes from other districts.
    We use data from Census data of 2011~\cite{census2011} to create population samples for districts.
    
    \item
    Mobility during lockdown:
    When lockdown interventions are imposed at different stages, the corresponding links are broken and mobilty is restricted inside different zones, along with a compliance factor. Severity of lockdown can be manyfold -- from containment zones, district lockdown and sealing of state border. See Table~\ref{tab:param_notation3}, Table~\ref{tab:param_notation4}, and Table~\ref{tab:mobility matrix3} in appendix for details. 
    
\end{itemize}


\subsection{Modeling the epidemic} \label{subsec:epid}
\begin{figure}[h]
    \centering
    \includegraphics[width=0.42\linewidth]{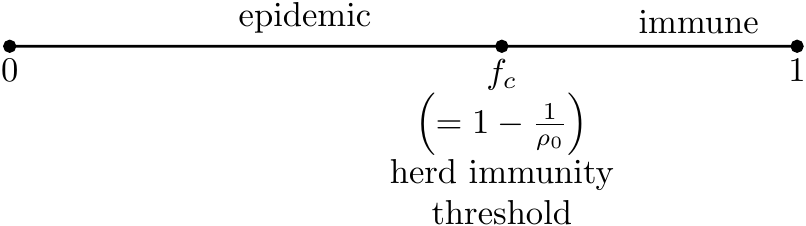}
    \caption{The epidemic phase diagram with the herd immunity threshold $f_c$ in terms of the basic reproduction number $\rho_0$.
    }
    \label{fig:R0}
\end{figure}

In the standard mathematical treatments of epidemics, there is a concept of the \textit{basic reproduction number} $\rho_0$ which is the expected number of cases generated by an infected one in a population where there is no immunity and all are susceptible to infection. This is not a fundamental constant, and depends on behavioral properties of the given population, and hence can differ across countries for the same disease. In fact, recent reports on dependence of weather suggest that the effective reproduction number $\rho$ for COVID-19 decreases by around $3\%$ per degree of increase of temperature beyond $25$C~\cite{xu2020weather}.
The generation and incubation times are rather more fundamental, as is the serial interval of infections.
In case a fraction $f$ attains immunity to the disease due to some reason, an infected individual will, on the average, infect $\rho = \rho_0(1-f)$. For the epidemic to die, needs to satisfy the condition $\rho < 1$, i.e., $f > 1-\frac{1}{\rho_0}$.  $f_c=1-\frac{1}{\rho_0}$ is known as the \textit{herd immunity threshold}. In this simplistic picture, this infection spreading is a \textit{branching process}, and is basically a \textit{percolation problem}~\cite{stauffer1994introduction}, in the language of statistical physics. There is a critical phenomenon associated with the phase transition, with the critical point at $f_c$, separating the epidemic phase ($f<f_c$) and the immune phase ($f>f_c$) (Fig.~\ref{fig:R0}).

While the early findings from COVID-19 cases in China suggested a value of $2.2-2.7$, the latest findings suggest a much higher average value of $\rho_0$ around $5.7$~\cite{sanche2020high}, which pushes the herd immunity threshold from around $60\%$ to above $80\%$. This implies that more than $80\%$ of the population needs to have immunity, naturally by recovering or through vaccination, for the infections to stop spreading further. 

The transmission of the virus is known to happen through various routes, mostly through exhaled droplets, but surfaces and fecal-oral contamination have also been reported.
The transmission are either through the symptomatic, pre-symptomatic, asymptomatic or environmental in nature. In what follows, we do not take the environment into account, explicitly. As in a macroscopic mathematical description of the epidemic~\cite{kermack1927contribution,aron1984seasonality}, we model the states of an individual in a population as the following: (i) Susceptible \textbf{S}, (ii) Exposed \textbf{E}, (iii) Infected \textbf{I}, (iv) Recovered \textbf{R} and (v) Dead \textbf{D}. We will denote the corresponding fractions of these states in the population as $S$, $E$, $I$, $R$ and $D$ respectively. This is similar to the SEIR model with an additional state \textbf{D} corresponding to the dead. The microscopic dynamics of the states of the model are described in the following.
\begin{figure}[h]
    \centering
    \includegraphics[width=0.60\linewidth]{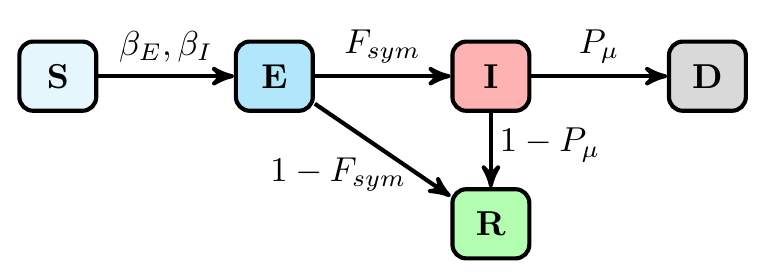}
    \caption{The basic SEIRD model: Susceptible $S$ can come in contact with the exposed $E$ or the infected $I$ to transition to exposed $E$ with rates $\beta_E$ and $\beta_I$ respectively. $F_{sym}$ of the total exposed agents \textbf{E} can mature into infected \textbf{I}. $1 - F_{sym}$ of the total exposed agents \textbf{E} may stay asymptomatic and eventually recover to state \textbf{R}. Infected agents may die to \textbf{D} with a probability $P_{\mu}$ or can recover to \textbf{R} by probability $1 - P_{\mu}$.
    }
    \label{fig:SEIRD}
\end{figure}

The initial state for the entire population is set to \textbf{S}. Epidemic spread is initialised with a few infected  agents $I_{init}$ which represent imported cases.
With each passing time-step, agents move and interact with other agents, and the mobility characteristics are dictated by the agent level features as described in Sec.~\ref{subsec:pop}.
The infected \textbf{I} coming in contact with \textbf{S} can convert them to \textbf{E}  with rate $\beta_I$. Once the state \textbf{E} is produced in the population, they in turn can convert the susceptible \textbf{S} into \textbf{E} with rate $\beta_E$. $F_{sym}$ of the total exposed agents \textbf{E} can mature into infected \textbf{I}. $1 - F_{sym}$ of the total exposed agents \textbf{E} may stay asymptomatic and eventually recover to state \textbf{R}. An early report from ICMR~\footnote{Indian Council of Medical Research, \texttt{https://www.icmr.gov.in/}, (2020), [accessed 20 April 2020]} claimed that the asymptomatic cases constituted 80\% of the total COVID cases~\cite{icmr_asymptomatic}. However, we have used  60\% asymptomatic cases in our epidemic modelling (refer to Table~\ref{tab:param_notation2} in Appendix for the value of the fraction $F_{sym}$). The infected agents may decease to \textbf{D} dictated by probability $P_{\mu}$ or \textbf{I} can eventually recover to \textbf{R} by probability $1 - P_{\mu}$.
Transitions to \textbf{I}, \textbf{R} and \textbf{D} are bounded by their corresponding time periods --
the time to infection $T_I$, recovery from infected $T_{RI}$, asymptomatic recovery from exposed $T_{RE}$, and mortality time $T_M$.
Lockdown intervention mechanism is also triggered which locks out districts depending upon the deployed strategy (for details refer to Sec.~\ref{subsec:lock}). This controls the contact process and hence the rapid spread of the infections. Fig.~\ref{fig:family} in Appendix shows an illustration of epidemic spread for \textit{Mumbai suburban} district at different temporal snapshots.

In order to keep track of the individuals in our agent based model, we carry out the microscopic simulation of the above dynamics using the Gillespie algorithm~\cite{gillespie1976general}.

\section{Results}
We use the model described in the previous section to simulate the behavior of agents in a population, with the epidemic process running on top of it, spreading through the process of contacts between individuals and spatially propagating due to mobility of exposed and infected individuals. 

We initiate our simulations with susceptible agents in the population, while we seed the epidemic process by assuming a few infected agents. The first quantities to calculate are the fractions $S$, $E$, $I$, $R$ and $D$ and their evolution over time. Using a model of their choice, the usual studies compute the infected fraction and try to match the numbers from reported data. In reality, reported data does not give us the correct number of infected population, specially for a socially complex, and populous country like India. This is because, it is practically impossible to test each individual and trace their contacts for possible further testing. Usually, the rate of testing is low and is unable to capture the real numbers. Moreover, there is a lag between testing and reporting of results, and this can be a major factor which makes tracing difficult. Trying to attempt to match reported number of cases with simulations will be a futile exercise.  

In our work, we do not attempt to match the number or fraction of infected individuals to real data. Our model, on the other hand, keeps information regarding each individual's movements in space and contact with others, enabling us to theoretically match the tracing and testing scenario, within the limits of practical error in real data.

\subsection{Contact Tracing}
For a disease like COVID-19, where the incubation period is long and the fraction of asymptomatic population is quite large, a standard method of relying on people's reporting accuracy to trace who they came in contact with, is insufficient. In that case, the tracing starts with an individual showing symptoms and its contacts can possibly be traced both forward and backward in time, the latter being more important because it can enable tracing the other possible paths the infection has branched into, and thus recursively  trace the parent of the transmission sub-tree. In this digital era of smartphones, contact tracing can be enhanced, not only in terms of the number of contacts but also in real time. The theoretically zero time lag between occurrence of a contact and it being reported is a matter of technology.  
\begin{figure}[h]
    \includegraphics[width=0.49\linewidth]{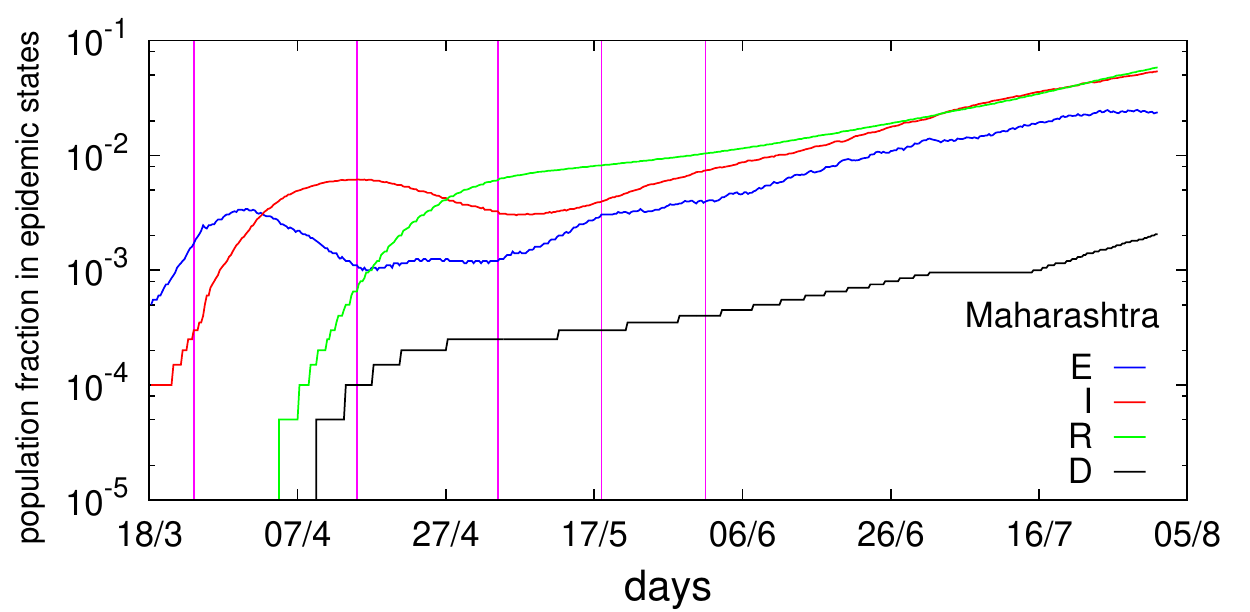}
    \includegraphics[width=0.49\linewidth]{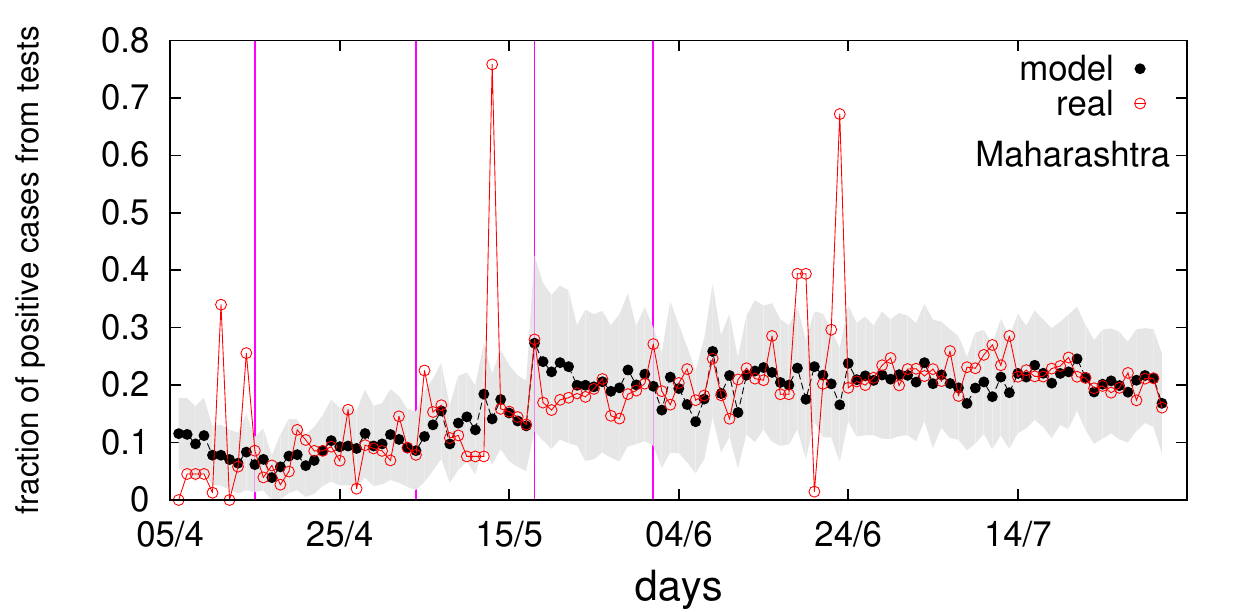}
    \includegraphics[width=0.49\linewidth]{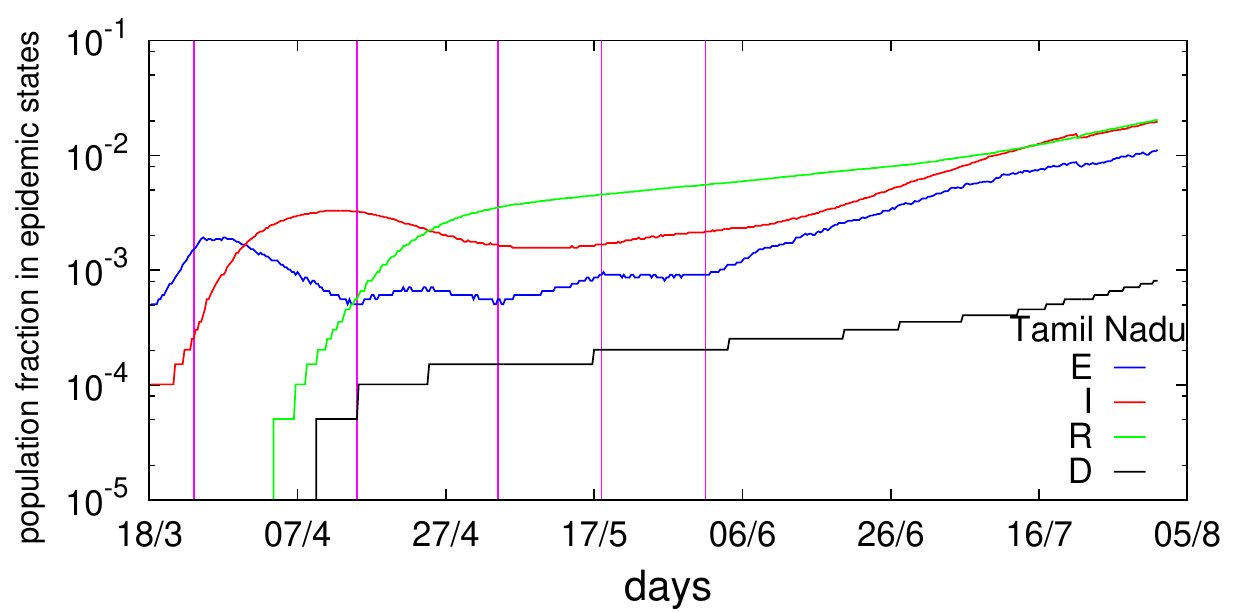}
    \includegraphics[width=0.49\linewidth]{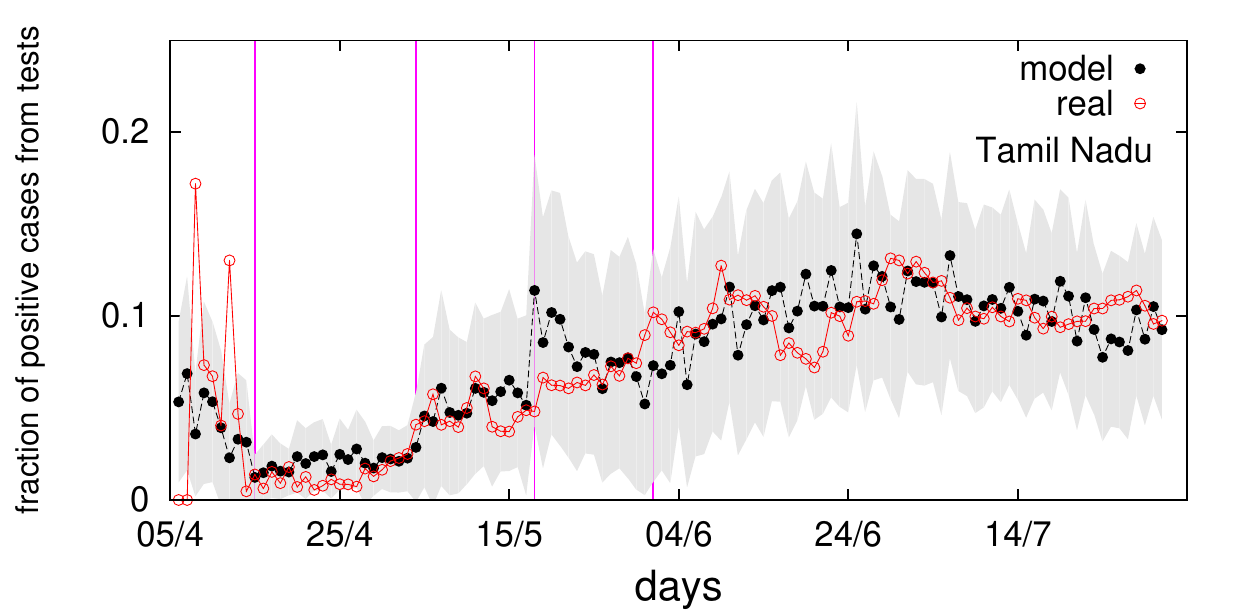}
    \caption{
    (Left Panel) The fractions $E$, $I$, $R$ and $D$ over time as observed in numerical simulations. (Right panel) The corresponding test positive rates compared with real data over time (right panels) for the states of Maharashtra and Tamil Nadu. The results are shown for 10 networks and 10 runs for each network for both the states. Maharashtra has 35 districts and Tamil Nadu has 32 districts (as per 2011 census) and our simulation results are shown for a sample population of 19944 agents and 19727 for Maharashtra and Tamil Nadu  respectively. The vertical lines correspond to the change in intervention (lockdown) rules in India (25/03, 15/04, 04/05, 18/05, 01/06).}
    \label{fig:lockdown_sim}
\end{figure}

\subsubsection{Testing rate}
The most important part of creating a microscopic agent based model with a variety of agent features, mobility behavior and detailed epidemic parameters, is to be able to correspond to the scenario where each agent can be traced spatio-temporally for its contacts. In real data, the \textit{testing rate} is defined as the number of individuals tested per unit population. Another important quantity of our interest is the \textit{test positive rate}, which is given by the number of individuals testing positive per unit tested population. In reality, individuals are rarely tested completely randomly because that is economically inefficient. Usual procedures include testing someone who reports a symptom, then test a list of its contacts, and maybe more -- secondary contacts, and so on. Often individuals who have been in vicinity of an area which reports a lot of infected cases, are also tested.


We use the real data~\cite{covid19indiaorg2020tracker} of testing and \textit{test positive rate} in a given state and try to fit that from our simulation results. In doing so, a set a parameters are used for fitting. 
When an agent gets infected, all the contacts of the agent for the last $T_{LB}$ days along with $N_{RT}$ number of random agents are tested for the infection. 
The real scenario corresponds to the infected agent being able to recollect (or through a tracing app) all other agents who came in contact in that span of time $T_{LB}$. $N_{RT}$ can be interpreted as an offset term taking care of the error and delay in tracing.
In Fig.~\ref{fig:lockdown_sim} we show the correspondence between the real data for the fraction of positive cases over time, and tried to match our simulation results by tuning several parameters related to mobility and lockdown compliance (see Table~\ref{tab:param_notation3} in Appendix for details).
We find the best fits for  $T_{LB}=5$ days and $N_{RT}$ is either 2 or 3 depending on the lockdown regime.

In our simulations, we constructed 10 different networks for the sample populations and simulated 10 runs of the epidemic process on each of them. The mean and standard deviations of the test positive rates are shown, along with the real data. In the Indian context, the strictness of the lockdown had varied through the timeline and thus the parameters have been altered accordingly (see Table~\ref{tab:param_notation3} in Appendix for details). These are also taken into consideration during the simulation. However, there are irregular spikes in the empirical data since the reporting is known to irregular, with data missing for few days and being accumulated for a particular day when they are actually available.

The corresponding density of different epidemic states in the population is also shown (left panels of Fig.~\ref{fig:lockdown_sim}) for the sample populations from Indian states of Maharashtra and Tamil Nadu. The growth of the infected state \textbf{I} is found to be slow, with a slow growth rate consistent what is observed in empirically reported data.

\subsection{Restricting movement using lockdown strategies}
\label{subsec:lock}

Containing infections being the main objective, the primary task is restricting the contacts of the population by arresting their movement at different scales. This leads to asking the question that what could be a viable strategy which is both operationally easy and functional, as well as effective in containment. Of course, the brute global lockdown can slow down the spreading but is not the best option for several reasons, mainly because certain economic sectors need to mobilize to facilitate the restart of the economy, which otherwise during the restriction period comes to a standstill.

In the following, we discuss a few different strategies that can be realized. We try to use the information about the bare fraction of infections in a given area and subsequently use the instantaneous reproduction number for the infections as triggers for deciding when to impose and relax restrictions in the different constituent districts of a given state. We also try a more sophisticated reinforcement learning based protocol to make the same decisions. In all the results that follow, we assume that no pharmaceutical interventions are being used, and the population is left to itself to acquire immunity through recovering from the infections. Our simulated time horizons are long for demonstration, as a consequence of the above assumption.

\subsubsection{Using fraction of infected cases}
\begin{figure}[h]
    \includegraphics[width=0.49\linewidth]{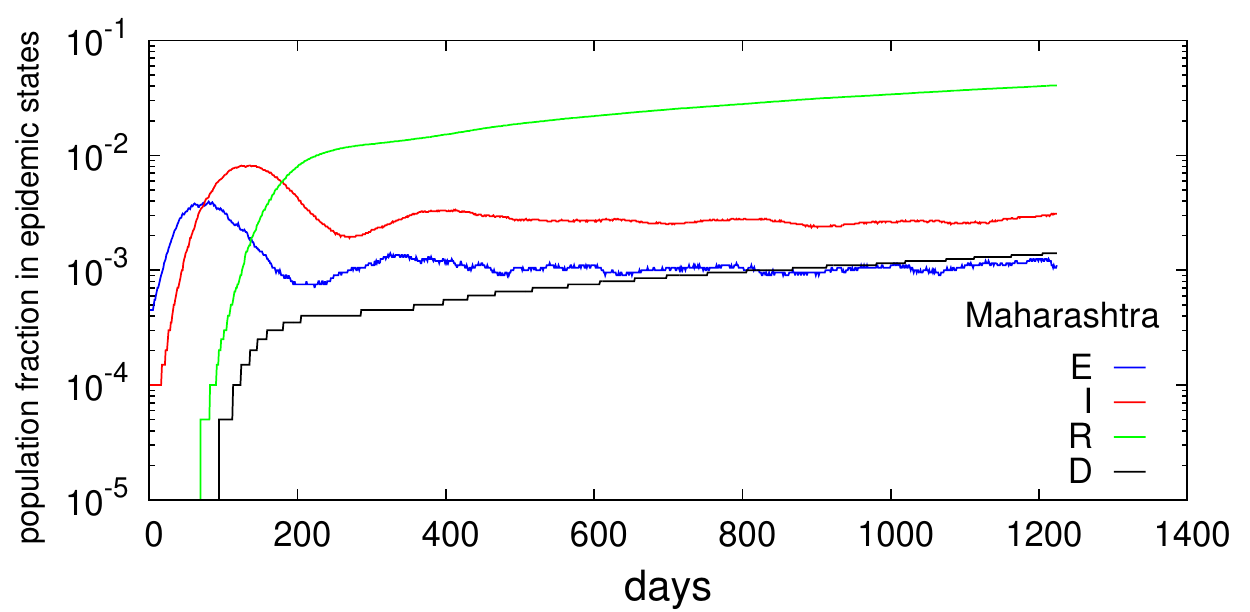}
    \includegraphics[width=0.49\linewidth]{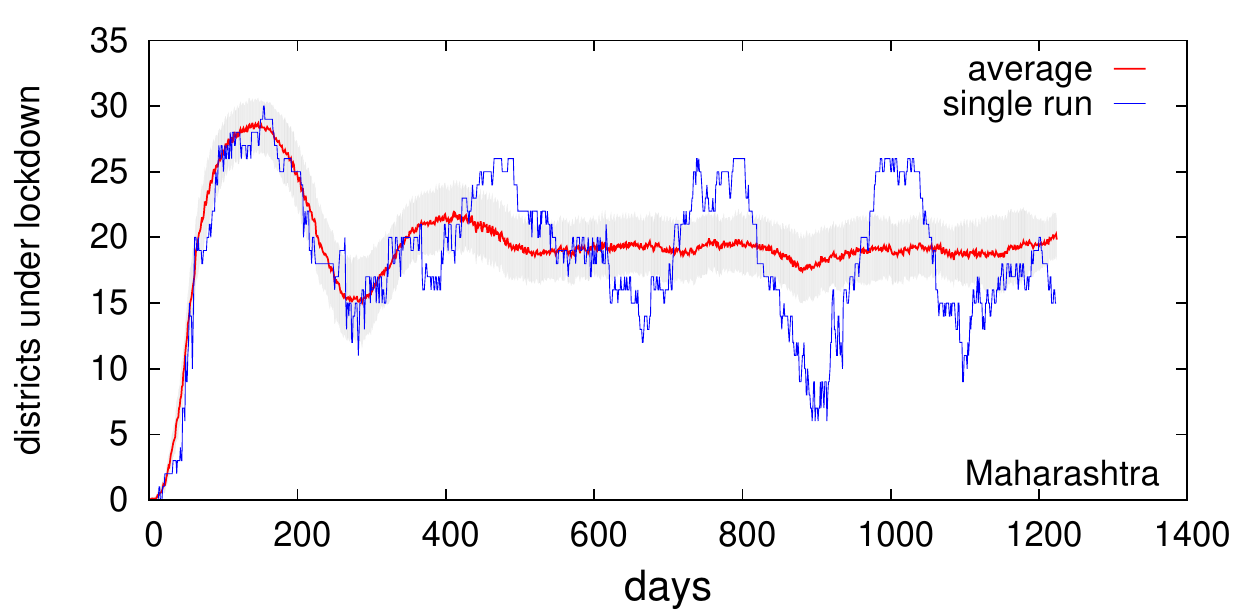}
    \caption{
    The simulation results for lockdown on districts depending on the fraction of infections. Lockdown interventions are activated when infections are greater than $I_u$ and deactivated when less than $I_d$. The result shown is the average for 50 runs (10 runs each on 5 networks ).
     (Left panel) The fractions $E$, $I$, $R$ and $D$ over time. (Right panel) The average number of districts under lockdown at a certain day  for the state of Maharashtra. The grey shaded region denotes one standard deviation. The averaged data is compared with a single run on a particular network realisation. Maharashtra has 35 districts (as per 2011 census) and our simulation results are shown for a sample population of $19943$ agents.}
    \label{fig:lockdown_sim_i}
\end{figure}

The natural trigger for imposing lockdown interventions will be by looking at the fraction of infections in a particular area. We use districts as the geographical entity where the fraction of infections is calculated and depending on thresholds $I_u$ and $I_d$, the interventions are turned on or off. In our simulations, shown in Fig.~\ref{fig:lockdown_sim_i}, we demonstrate the case when the intervention switch is $I_u = 1.75 \times 10^{-3}$, i.e. 0.175\%, i.e. interventions are effective when $I(t) > I_u$ and relaxed when $I(t) < I_d$ with $I_d = 1.3125 \times 10^{-3}$, i.e. $0.13125\%$ (See Table~\ref{tab:param_notation5} in Appendix). In the figure, we also demonstrate how the number of districts under lockdown behave for a single run on a particular network, showing considerable fluctuation around the average value. 

\subsubsection{Using reproduction rate}
\begin{figure}[h]
    \includegraphics[width=0.49\linewidth]{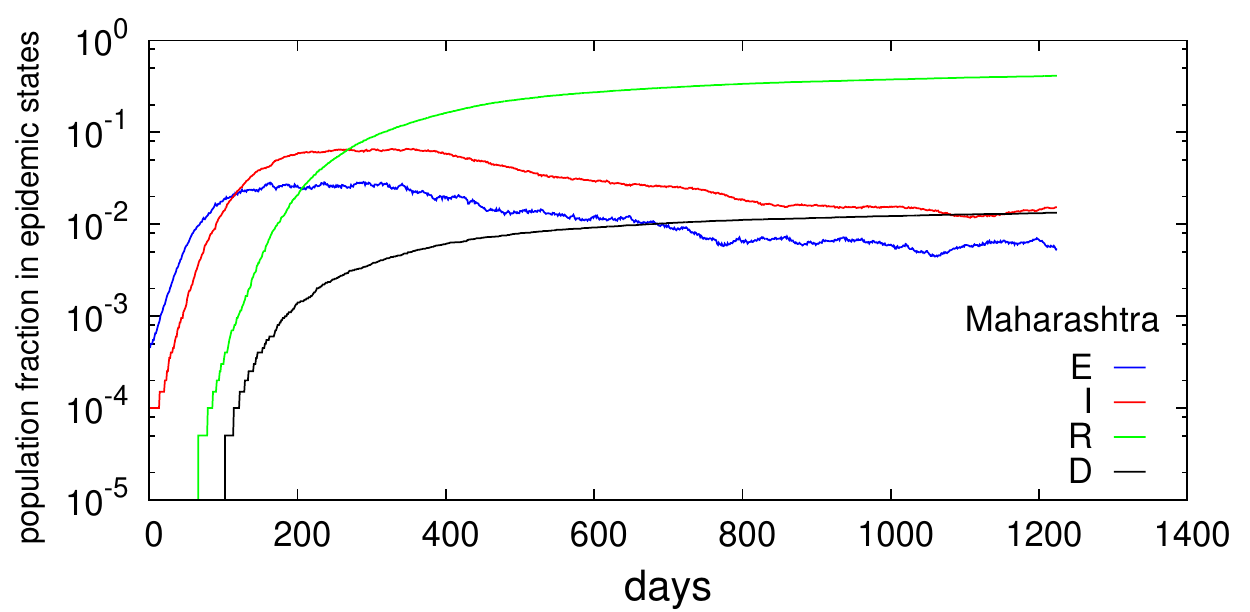}
    \includegraphics[width=0.49\linewidth]{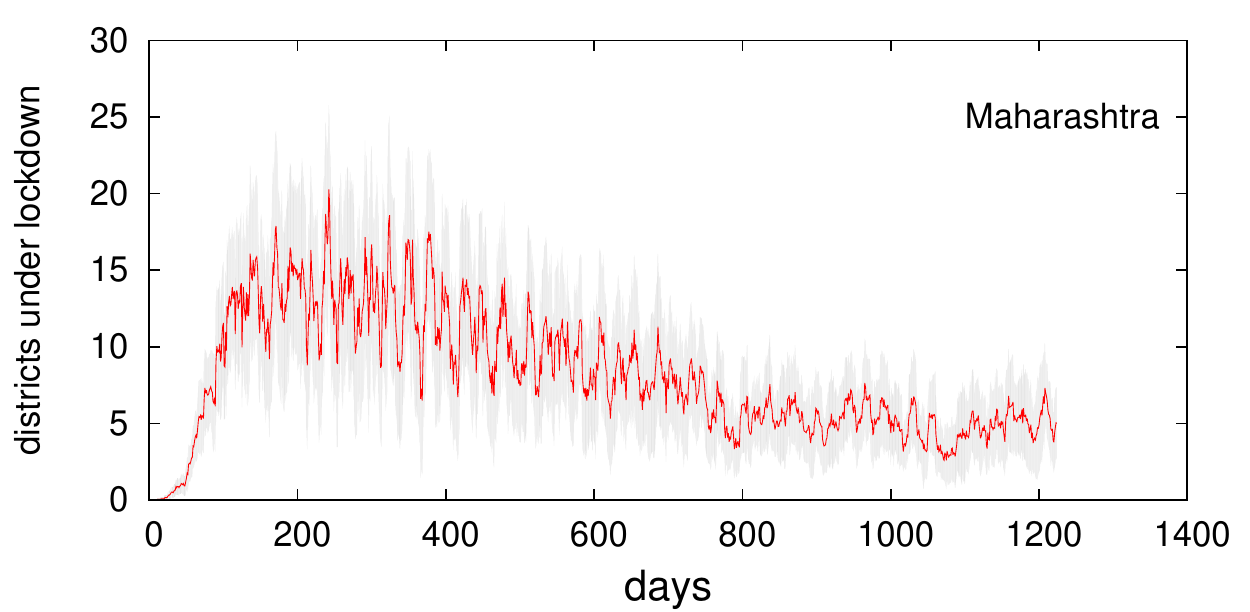}
    \caption{
    The simulation results for lockdown on districts depending on the reproduction rate $\rho(t)$. 
     Lockdown interventions are activated when reproduction rate $\rho(t)$ is greater than 1 and deactivated when less than 0.5.
     The results are averaged over 25 runs (5 runs each on 5 networks).
     (Left panel) The fractions $E$, $I$, $R$ and $D$ over time.  
     (Right panel) The average number of districts under lockdown at a certain day for the state of Maharashtra. The grey shaded region denotes one standard deviation. Maharashtra has 35 districts (as per 2011 census) and our simulation results are shown for a sample population of 19943 agents.}
    \label{fig:lockdown_sim_r}
\end{figure}

Compared to using the fraction of infected cases, one can imagine a rather sophisticated way of designing a lockdown trigger by using the instantaneous effective reproduction number $\rho(t)$. The values of $\rho(t)$ can be calculated at the level of different states of India (following the recipe of Ref.~\cite{bettencourt2008real} and is shown in Fig.~\ref{fig:rhot_emp} in Appendix). In our model simulations, we compute the values of $\rho(t)$ for each district. We prescribe a rule where a district goes under lockdown if the $\rho(t)$ exceeds a particular threshold, $\rho(t) > \rho_u$ and is relaxed when is below another low threshold, $\rho(t) < \rho_d$.
In Fig.~\ref{fig:lockdown_sim_r}, we show the case when the interventions are in place while $\rho_u = 1$ and relaxed when $\rho_d = 0.5$ (See Table~\ref{tab:param_notation5} in Appendix).

\subsubsection{Using reinforcement learning}

Critical factors in decision making while imposing lockdown include infections, recoveries, deaths and economic loss. In previous sub-sections, we focused on infection and recovery as triggers whereas death and economic impact were not considered. To address this, we used reinforcement learning (RL). Reinforcement Learning is a technique in which an agent learns and improves using its experience and performs actions while interacting with environment with a goal to maximize reward. Observable characteristics of the environment is referred to as state and its representation is given as context to the agent.

\begin{figure}[h]
    \includegraphics[width=0.49\linewidth]{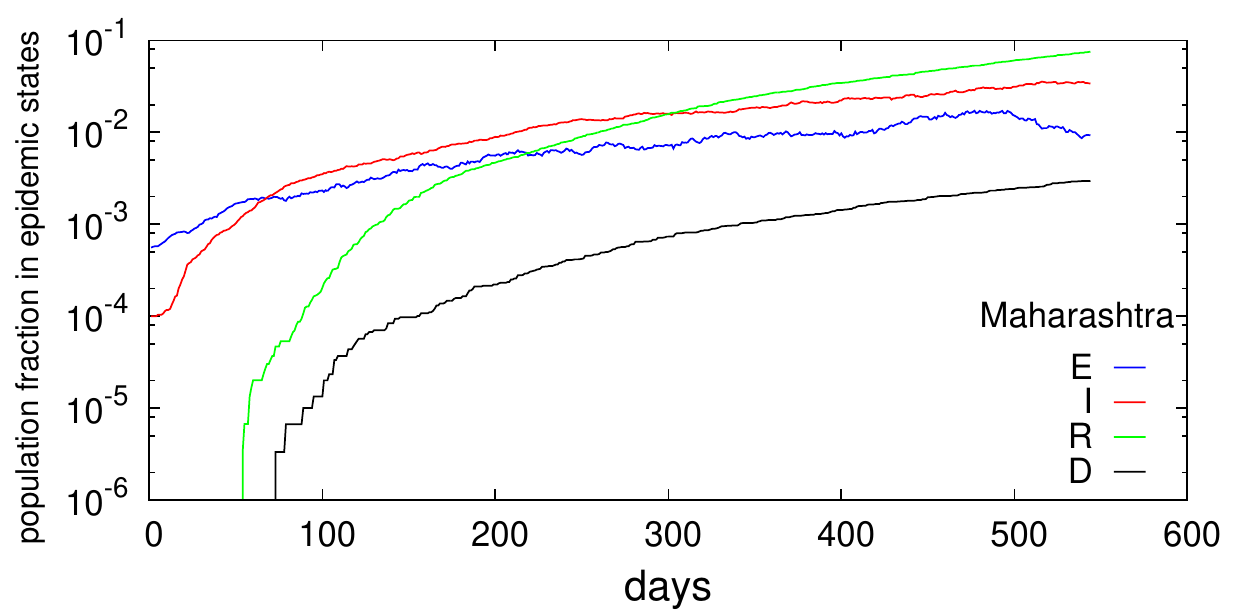}
    \includegraphics[width=0.49\linewidth]{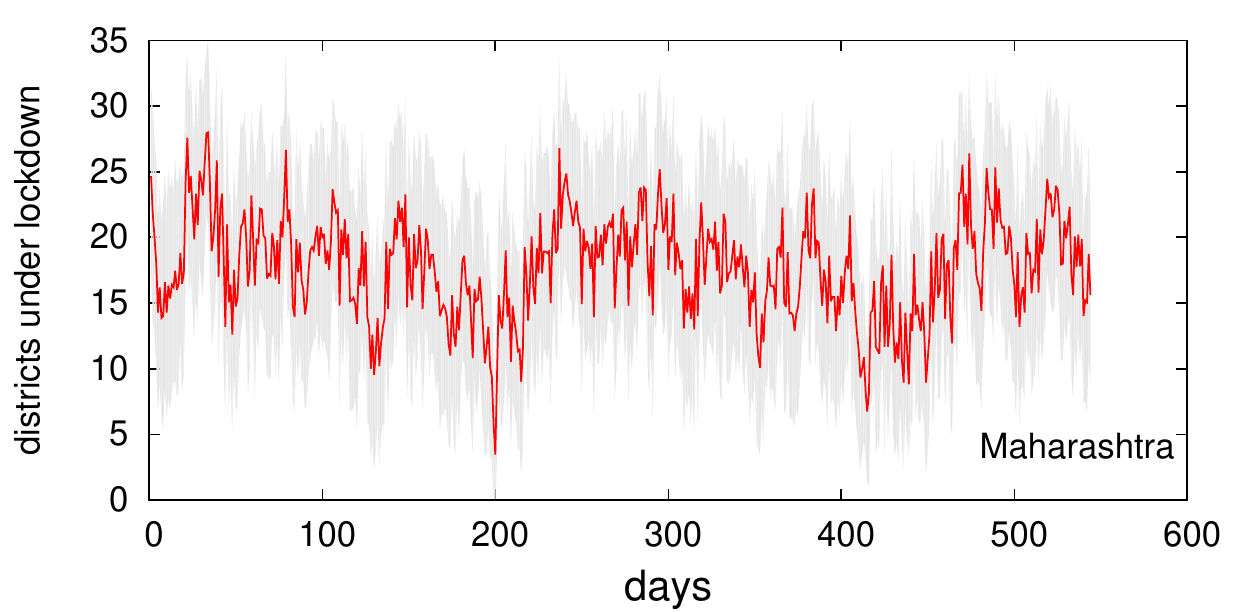}
    \caption{
    The simulation results for lockdown on districts as proposed by reinforcement learning. The results are shown for 5 networks and 3 runs each of 100 episodes for each network.
      (Left panel) The fractions $E$, $I$, $R$ and $D$ over time.
      (Right panel) The average number of districts under lockdown at a certain day for the state of Maharashtra.The grey shaded region denotes one standard deviation.  Maharashtra has 35 districts (as per 2011 census) and our simulation results are shown for a sample population of 19943 agents.}
    \label{fig:lockdown_sim_rl}
\end{figure}
We trained a Deep Q Network (DQN) \cite{mnih2015humanlevel} agent to learn an optimal intervention strategy. State representation presented to the RL agent in order to choose an action (imposing lockdown or not for every district) consisted of district-wise population, infections, recoveries, deaths, infection rate, recovery rate and the death rate at every time-step. An action with a higher Q value is chosen greedily by the RL agent during exploitation phase. Whereas, in the exploration phase, actions are chosen randomly. Rewards are calculated on the basis of number of districts under lockdown, infection count and death count for every district. Training was carried out using 2 hidden layers with learning rate $= 0.001$ and momentum $= 0.8$ for $10$ epochs (similar to Ref~\cite{khadilkar2020optimising}) using the mean squared error as the loss function. Fig.~\ref{fig:lockdown_sim_rl} shows the essentials of epidemic spread using trained DQN agent for the state of Maharashtra.

\section{Discussions}
The COVID-19 pandemic has become one of the threats to humanity, by spreading rapidly, causing life threatening clinical complications by itself and somewhat fatal for individuals with other co-morbidities~\cite{clark2020global}. With less information about the epidemic in the first few months, healthcare systems were overwhelmed with critical cases. With time, some reliable scientific information  has helped healthcare systems across different countries to cope with the pandemic. However, the number of cases have multiplied by then. Researchers have explored various classical and improvised modeling techniques to understand and predict the rising number of cases, with little success.

Modeling epidemics in real populations for studying various possible scenarios and enable prediction is an extremely challenging task. While basic mathematical models established over time provide the necessary framework for realizing the population and the processes of spreading, relating to real data is tricky. This is mainly because the data captured through surveys, reporting and testing are not at par with the actual information on the infections in the population. For new pathogens, whose transmission and infectious behavior (e.g., symptoms, detection, etc.) are not well established or known, the ambiguity contributes to the error in estimating the actual number of infections and its temporal characteristics. In most countries, the test protocol and reporting of results are quite erratic and thus the gap between the actual infections and the measured, is beyond the scope of modeling. Under these circumstances, almost all of the modeling efforts for COVID-19 are thus, not accurately able to estimate or predict the actual number of infections. In our study, we do not attempt to match the actual number of reported cases as well.

In our approach, we have constructed a framework that takes into account the spatio-temporal complexity of the population and its contact characteristics. Our study, for the case of India, makes use of the available Census data for extracting social, demographics and the spatial population features, along with hypothesis driven modeling of spatially hierarchical organization of population movement and contact. This enables us to monitor the dynamics of the epidemic through contact at various spatial scales. 
Using our microscopic, agent based model of the COVID-19 epidemic for Indian population, we demonstrate our results for sample populations at the level of `states', the basic federal administrative units of India.
However, the framework can be used to focus small populations in municipalities. On the other hand, the model can be scaled up to the level of the country using appropriate modifications due to realization of multiple `states'.
The simple, toy model approaches help in understanding the gross behavior of the epidemic and its states, but mostly fail to answer detailed questions related to the population, demographic and spatial features. Although our detailed model contains many parameters, most of them can be tuned and validated with real data, whenever and wherever they are available. In that way, our approach provides a possibility of answering several fine grained questions.

We demonstrated how the framework can be used to compute the fraction of cases tested positive in a sample population whose contact behavior is regulated by the restrictions in movement due to imposed lockdowns.
The proposed framework also allows to test hypothetical lockdown protocols to study the time evolution of densities of epidemic states. In fact, we demonstrated the use of three different lockdown strategies for a particular state of India. To study this, we utilized the fraction of infections $I(t)$, the reproduction rate $\rho(t)$ and also reinforcement learning to decide
which of the districts should come under lockdown restrictions.
Our results show that these three methods can be utilized to dynamically decide lockdown in various districts and thereby restrict  the spread of infections. However, from the point of view of socio-economic impact of lockdowns, one can compare these protocols as well as new ones to decide which can be deployed in a real population. 
Regarding the feasibility as well as practicality of switching lockdowns on and off at a rapid rate, our results merely demostrate a possibility, which can be modified according to administrative capabilities. In a realistic scenario, lockdowns may not be implemented with a possibility of changing state at the scale of a single day.
In addition, our model formulation is flexible in answering feature related queries, as well as adapting to new features which may be added for the population, spatial information as well as disease characteristics.






 





\bibliography{sample}

\clearpage

\appendix

\section*{Appendix}

\subsection*{Transition rates in the epidemic model}
Rate parameters $\alpha$, $\delta$, $\gamma$, $\mu$ are respectively used in 
 Algorithm~\ref{alg:EpidemicModel} and Algorithm~\ref{alg:GetNwParams}
for calculating the total infection, recovery from exposure, recovery from infection and death rates respectively. The corresponding time periods are already defined as $T_I$, $T_{ER}$, $T_{IR}$ and $T_M$ respectively in Sec.~\ref{subsec:epid}.
These rates are used to decide the transition, whereas parameters $F_{sym}$ and $P_{\mu}$ are used for selecting eligible nodes for transition.

\subsection*{Tuning and changing parameters during the lockdown periods}
Isolation compliance for infected $C_{I}$ has been kept a constant throughout the simulation period as symptomatic agents have been following the same norms throughout. As testing rate has increased gradually in different phases of lockdowns, isolation compliance for exposed agents $C_{E}$ increases with the increasing phases of lockdowns. Border seal compliance $C_{BS}$ decreases with lockdown phases as border seal policies become less stringent with passing time and inter-district mobility is allowed with varying degrees in different phases of the lockdown. The probability of returning to home $C_{H}$ is least when there is no lockdown as people are free to move and stay away from their homes when there is no intervention. It increases in the first phase of the lockdown due to curbing of mobility and then decreases in the subsequent phases (constant in last three phases) as restrictions are lifted gradually to some extent. The probability of removing an edge $P_{re}$ is the probability of breaking an existing edge of an agent when it moves to the same or a different zone. This probability is very low when there is no lockdown and is high during lockdowns as agents tend to break contacts and keep it to minimum during lockdowns. The probability of creating an edge $P_{ce}$ is kept high when there is no lockdown as agents move and make contacts freely. However, this probability is very low in the first phase of lockdown due to strict social distancing norms and increases with small values in the subsequent phases as restrictions on mobility are lifted gradually. As inter-family interactions are limited during lockdowns, the probability of removing edges between families is quite high during lockdowns and decreases with lockdown phases as families start moving and interacting with each other with more liberty. The probability of moving to a different district $M_{DD}$ is comparatively lower during lockdowns as agents move to other districts only if it is very urgent. Refer to Table~\ref{tab:param_notation3} for the values of these parameters during different lockdown phases.


\newpage

\begin{table}[h]
  \centering
  \caption{Parameters of network part of the model, with notations for the variables and parameters}
  \label{tab:param_notation1}
   \resizebox{\textwidth}{!}{ 
  \begin{tabular}{|l|l|l|}
  \hline 
  \textbf{Network Parameter} & \textbf{Notation} & \textbf{Remarks}\\ \hline
  Population network & $N$ & Hierarchical network formed using Algorithm 1\\ \hline
    Household size distribution for a district & $P_{household}$ & India Census 2011 data~\cite{census2011} \\ \hline
    Age distribution for a district & $P_{age}$ & India Census 2011 data~\cite{census2011}\\ \hline
    Gender distribution for a district & $P_{gender}$ &  India Census 2011 data~\cite{census2011}\\ \hline
    Age-wise probability of employment
    & $P_{emp}$ & India Census 2011 data~\cite{census2011}\\ \hline
    Fraction of essential workers & $F_{essential}$ & 0.01\\ \hline
    Number of zones & $N_{zones}$ & 25\\ \hline
    Minimum number of families in a district & $Min_{families}$ & Backtracked from state's population size to get 20k nodes for each network\\ \hline
    Name of the district & $district\_name$ & India Census 2011 data~\cite{census2011}\\ \hline
    Name of the state & $state\_name$ & India Census 2011 data~\cite{census2011}\\ \hline
    District's Latitude-Longitude & $district\_latlong$ & Mapped from GeoPy \footnote{https://geopy.readthedocs.io/en/stable/\#}\\ \hline
    State's Latitude-Longitude & $state\_latlong$ & Mapped from GeoPy \\ \hline
    Epidemic state of a node & $ep\_st$ & Susceptible(S), Exposed(E), Infected(I), Recovered(R), Dead(D) \\ \hline
    Last update time of a node & $time$ & Day(relative to start of the simulation) when the state of agent changed last \\ \hline
    Node's free mobility flag & $is\_free$ & Parameter to control node's movement \\ \hline 
    

\end{tabular}
}
\end{table}

\begin{table}[h]
  \centering
  \caption{Parameters of epidemic part of the model and notations for variables and parameters}
  \label{tab:param_notation2}
   \resizebox{\textwidth}{!}{ 
  \begin{tabular}{|l|l| l|}
  \hline 
    \textbf{Epidemic Model Parameter} & \textbf{Notation} & \textbf{Remarks}\\ \hline
    Initial infections & $I_{init}$ & 2 \\ \hline
    Contact rate for exposed population & $\beta_{E}$ & Sampled on daily basis for each node from Gamma distribution, shape=8.0, scale=0.6, constant=1.2~\cite{ferretti2020quantifying}\\ \hline
    Contact rate for infected population & $\beta_{I}$ &  Sampled on daily bases for each node from Gamma distribution, shape=18.0, scale=0.4, constant=0.85~\cite{ferretti2020quantifying}\\ \hline
    Adjusted case fatality ratio & $P_{\mu}$ & Specific to each Indian state (Code \footnote{https://cmmid.github.io/topics/covid19/global\_cfr\_estimates.html}, Data \footnote{https://www.ecdc.europa.eu/en/publications-data/download-todays-data-geographic-distribution-covid-19-cases-worldwide} \cite{Golding2020.07.07.20148460})\\  \hline
    Fraction of symptomatic nodes & $F_{sym}$ & 0.4 ~\cite{icmr_asymptomatic}\\  \hline
    Incubation time & $T_{I}$ & Gamma distribution, shape=5.8, scale=0.95~\cite{linton2020incubation,lauer2020incubation}\\ \hline
    Recovery time from infection & $T_{RI}$ & Gamma distribution, shape=1400.0, scale=0.01 ~\cite{WHOrt} \\ \hline
    Recovery time from exposure & $T_{RE}$ & Gamma distribution, shape=25.0, scale=0.9 ~\cite{WHOrt}\\ \hline
    Mortality time from infection & $T_{M}$ & 13 days \cite{wilson2020} \\  \hline
    Maximum simulation days & $T_{max}$  & 136\\ \hline 
\end{tabular}
}
\end{table}

\begin{table}
  \centering
  \caption{Parameters of model and notations for variables and parameters: Fixed day intervention}
  \label{tab:param_notation3}
   \resizebox{\textwidth}{!}{ 
  \begin{tabular}{|l|l|l|l|l|l|l|l|}
  \hline 
    \textbf{Epidemic Model Parameter} & \textbf{Notation} & \textbf{No Intervention} & \textbf{Intervention 1} & \textbf{Intervention 2}
    & \textbf{Intervention 3} & \textbf{Intervention 4} & \textbf{Intervention 5}\\ \hline
    Intervention application dates (DD/MM) & $Dt_{Int}$ & 18/03 & 25/03 & 15/04 & 04/05 & 18/05 & 01/06 \\ \hline
    Isolation compliance for infected & $C_{I}$ & 0.7 & 0.7 & 0.7 & 0.7 & 0.7 & 0.7 \\ \hline
    Isolation compliance for exposed & $C_{E}$ & 0.3 & 0.35 & 0.4 & 0.5 & 0.55 & 0.6 \\ \hline
    Border seal compliance & $C_{BS}$ & 0 & 0.7 & 0.65 & 0.6 & 0.5 & 0.4 \\ \hline
    Probability of returning to home at night & $C_{H}$ & 0.8 & 0.9 & 0.875 & 0.85 & 0.85 & 0.85 \\ \hline
    Probability of removing an edge when an agent moves & $P_{re}$ & 0.2 & 0.7 & 0.7 & 0.7 & 0.7 & 0.7 \\ \hline
    Probability of creating an edge when an agent moves & $P_{ce}$ & 0.85 & 0.4 & 0.45 & 0.5 & 0.525 & 0.55 \\ \hline
    Probability of removing edges between families & $P_{fe}$ & 0 & 0.8 & 0.7 & 0.6 & 0.5 & 0.4\\ \hline
    Inter-district mobility matrix & $M_{DD}$ & 0.3 & 0.2 & 0.2 & 0.2 & 0.2 & 0.2 \\ \hline
    Look back period for contact tracing (in days) & $T_{LB}$ & 5 & 5 & 5 & 5 & 5 & 5 \\ \hline
    Random testing for contact tracing & $N_{RT}$ & 2 & 2 & 3 & 3 & 0 & 0 \\  \hline

\end{tabular}
}
\end{table}

\begin{table}
  \centering
  \caption{Parameters of model and notations for variables and parameters: Interventions based on (a) Number of infections in a district I(t), (b) Effective Reproduction Number $\rho(t)$, and (c) Reinforcement Learning.}
  \label{tab:param_notation4}
  \begin{tabular}{|l|l|l|l|}
  \hline 
    \textbf{Epidemic Model Parameter} & \textbf{Notation} & \textbf{No Intervention} & \textbf{Intervention} \\ \hline
    Isolation compliance for infected & $C_{I}$ & 0.7 & 0.9 \\ \hline
    Isolation compliance for exposed & $C_{E}$ & 0.3 & 0.9 \\ \hline

    Border seal compliance & $C_{BS}$ & 0 & 0.9 \\ \hline
    Probability of returning to home at night & $C_{H}$ & 0.8 & 0.95 \\ \hline
    Probability of removing an edge when an agent moves & $P_{re}$ & 0.2 & 0.7 \\ \hline
    Probability of creating an edge when an agent moves & $P_{ce}$ & 0.85 & 0.5 \\ \hline
    Probability of removing edges between families & $P_{fe}$ & 0 & 0.8 \\ \hline
    Inter-district mobility matrix & $M_{DD}$ & 0.3 & 0.2  \\ \hline
    Random testing for contact tracing & $N_{RT}$ & 2 & 5  \\  \hline
    Maximum simulation days for I(t) and $\rho(t)$ & $T_{max}$ & \multicolumn{2}{c|}{306}  \\ \hline
    Maximum simulation days for RL & $T_{max}$ & \multicolumn{2}{c|}{136}  \\ \hline
\end{tabular}
\end{table}

\begin{table}
  \centering
  \caption{Threshold for intervention}
  \label{tab:param_notation5}
  \begin{tabular}{|l|l|l|}
  \hline 
    \textbf{Epidemic Model Parameter} & \textbf{Upper threshold} & \textbf{Lower threshold} \\ \hline
    I(t) & $I_u=  0.00175$ & $I_d = 0.00131$  \\ \hline
    $\rho(t)$  & $\rho_u = 1$ & $\rho_d=0.5$  \\ \hline
\end{tabular}
\end{table}


\begin{table}
  \centering
  \caption{Mobility Matrix ($M_{ZZ}$) for weekdays in the absence of intervention (lockdown) for all source zones}
  \label{tab:mobility matrix1}
  \begin{tabular}{|l|l|l|l|l|l|l|}
  \hline 
    \multirow{2}{*}{\textbf{Age group}} & \multirow{2}{*}{\textbf{Time bin}} & \multicolumn{5}{c|}{\textbf{Destination zone}} \\ \cline{3-7}
    &  & \textbf{RR} & \textbf{EE} & \textbf{EPB} & \textbf{EPV} & \textbf{NM} \\ \hline
    \multirow{4}{*}{Age Bracket 1} & Morning & 0.4 & 0.45 & 0.01 & 0.04 & 0.1 \\ \cline{2-7}
    & Afternoon & 0.4 & 0.45 & 0.01 & 0.04 & 0.1 \\ \cline{2-7}
    & Evening & 0.47 & 0.34 & 0.07 & 0.02 & 0.1 \\ \cline{2-7}
    & Night & 0.46 & 0.29 & 0.13 & 0.02 & 0.1 \\ \hline
    
    \multirow{4}{*}{Age Bracket 2} & Morning & 0.05 & 0.5 & 0.25 & 0.1 & 0.1 \\ \cline{2-7}
    & Afternoon & 0.2 & 0.45 & 0.15 & 0.1 & 0.1 \\ \cline{2-7}
    & Evening & 0.225 & 0.375 & 0.175 & 0.125 & 0.1 \\ \cline{2-7}
    & Night & 0.29 & 0.13 & 0.24 & 0.24 & 0.1 \\ \hline
    
    \multirow{4}{*}{Age Bracket 3} & Morning & 0.2 & 0.6 & 0.05 & 0.05 & 0.1 \\ \cline{2-7}
    & Afternoon & 0.25 & 0.55 & 0.05 & 0.05 & 0.1 \\ \cline{2-7}
    & Evening & 0.3 & 0.5 & 0.05 & 0.05 & 0.1 \\ \cline{2-7}
    & Night & 0.55 & 0.25 & 0.05 & 0.05 & 0.1 \\ \hline

\end{tabular}
\end{table}

\begin{table}
  \centering
  \caption{Mobility Matrix ($M_{ZZ}$) for weekends in the absence of intervention (lockdown) for all age brackets}
  \label{tab:mobility matrix2}
  \begin{tabular}{|l|l|l|l|l|l|l|}
  \hline 
    \multirow{2}{*}{\textbf{Source zone}} & \multirow{2}{*}{\textbf{Time bin}} & \multicolumn{5}{c|}{\textbf{Destination zone}} \\ \cline{3-7}
    &  & \textbf{RR} & \textbf{EE} & \textbf{EPB} & \textbf{EPV} & \textbf{NM} \\ \hline
    \multirow{4}{*}{\textbf{RR}} & Morning & 0.22 & 0.52 & 0.1 & 0.06 & 0.1 \\ \cline{2-7}
    & Afternoon & 0.29 & 0.48 & 0.07 & 0.06 & 0.1 \\ \cline{2-7}
    & Evening & 0.33 & 0.41 & 0.1 & 0.06 & 0.1 \\ \cline{2-7}
    & Night & 0.43 & 0.19 & 0.16 & 0.12 & 0.1 \\ \hline
    
    \multirow{4}{*}{\textbf{EE}} & Morning & 0.22 & 0.66 & 0.01 & 0.01 & 0.1 \\ \cline{2-7}
    & Afternoon & 0.29 & 0.59 & 0.01 & 0.01 & 0.1 \\ \cline{2-7}
    & Evening & 0.33 & 0.55 & 0.01 & 0.01 & 0.1 \\ \cline{2-7}
    & Night & 0.43 & 0.45 & 0.01 & 0.01 & 0.1 \\ \hline
    
    \multirow{4}{*}{\textbf{EPB}} & Morning & 0.225 & 0.05 & 0.62 & 0.005 & 0.1 \\ \cline{2-7}
    & Afternoon & 0.295 & 0.05 & 0.55 & 0.005 & 0.1 \\ \cline{2-7}
    & Evening & 0.335 & 0.41 & 0.15 & 0.005 & 0.1 \\ \cline{2-7}
    & Night & 0.655 & 0.19 & 0.05 & 0.005 & 0.1 \\ \hline
    
    \multirow{4}{*}{\textbf{EPV}} & Morning & 0.225 & 0.05 & 0.005 & 0.62 & 0.1 \\ \cline{2-7}
    & Afternoon & 0.295 & 0.05 & 0.005 & 0.55 & 0.1 \\ \cline{2-7}
    & Evening & 0.335 & 0.41 & 0.005 & 0.15 & 0.1 \\ \cline{2-7}
    & Night & 0.655 & 0.19 & 0.005 & 0.05 & 0.1 \\ \hline

\end{tabular}
\end{table}

\begin{table}
  \centering
  \caption{Mobility Matrix ($M_{ZZ}$) during intervention (lockdown) for all zones and all age brackets.}
  \label{tab:mobility matrix3}
  \begin{tabular}{|l|l|l|l|l|l|}
  \hline 
    \multirow{2}{*}{\textbf{Time bin}} & \multicolumn{5}{c|}{\textbf{Destination zone}} \\ \cline{2-6}
    & \textbf{RR} & \textbf{EE} & \textbf{EPB} & \textbf{EPV} & \textbf{NM} \\ \hline
     Morning & 0.3 & 0.001 & 0.02 & 0.01 & 0.669 \\ \hline
    Afternoon & 0.1 & 0.001 & 0.02 & 0.01 & 0.869 \\ \hline
    Evening & 0.3 & 0.001 & 0.02 & 0.01 & 0.669 \\ \hline
    Night & 0.5 & 0.001 & 0.02 & 0.01 & 0.469 \\ \hline
    
\end{tabular}
\end{table}

\begin{algorithm}[H]
\caption{\texttt{Network Model}\\
$create\_zones\_grid(.)$: Assigns random zones to the grid within a district. \\
$get\_distributions(.)$: Returns the distribution of household size, age, gender and employment according to age for each district from census data. \\ 
$get\_complete\_graph(.)$: Create complete graph according to $P_{household}$. \\ 
$assign\_attributes(.)$: Assigns state, district, zone, grid and global (district) coordinates, age bucket, employment, gender, essential traveller flag (1 = essential, 0 = non-essential), epidemic state, last update time (used in epidemic modeling), mobility flag (1 = movable, 0 = non-movable) attributes to each node of a complete family graph.\\
$connect\_families(.)$: Connect families based on preferential attachment rule. \\
$get\_gravity\_attributes(.)$: Returns district's size and latitude-longitude. \\
$gravity\_law(.)$: Returns edges based on formula:
$\frac{size\_district_{i}\times size\_district_{j}}{d_{H}(latlong\_district_{i}, latlong\_district_{j})^2}$; where $d_{H}$ is haversine distance. \\
$get\_disjoint\_union(.)$: Returns a disjoint union graph of all districts without edges between them. \\
$sample\_nodes(.)$: Returns sampled nodes from two input district graphs based on the edges (calculated from gavity law) between them (with replacement).  \\
$connect\_edges(.)$: Connect the sampled nodes in the disjoint union graph. 
} 
\label{alg:CreateNetwork}
\begin{algorithmic}[1]
\Procedure{$create\_network$}{$Min_{families}, F_{essential}, N_{zones}, state\_name, district\_name, district\_latlong, ep\_st, time,  $ \par \hskip\algorithmicindent \hskip\algorithmicindent $ is\_free$}
\State $position\_zone\_mapping$ = $create\_zones\_grid(N_{zones})$
\State $P_{household},P_{age},P_{gender},P_{emp}$ =  $get\_distributions
(Min_{families})$
\For{$complete\_graph$ in $get\_complete\_graphs(P_{household})$}
    \State $families$ = $assign\_attributes(complete\_graph, P_{household},P_{age},P_{gender},P_{emp}, F_{essential}, state\_name, district\_name,  $ \par \hskip\algorithmicindent$ district\_latlong,  position\_zone\_mapping, ep\_st, time, is\_free)$
\EndFor 
\State $districts$ = $connect\_families(families)$ 
\State $state\_graph$ = $connect\_districts(districts)$ \\
\Return $state\_graph$
\EndProcedure
\end{algorithmic}

\label{alg:CreateComponent}
\begin{algorithmic}[1]
\Procedure{$connect\_districts$}{$districts$}
\For{$district_{i}$ in $districts$}
    \For{$district_{j}$ in $districts$}
        \State $size_{i}, size_{j}, latlong_{i},latlong_{j}$ = $get\_gravity\_attributes (district_{i},district_{j})$ 
        \State $edges\_btw\_districts_{i,j}$ = $gravity\_law(size_{i}, size_{j}, latlong_{i},latlong_{j})$
\EndFor
\EndFor
\State $disjoint\_union\_graph$ = $get\_disjoint\_union(districts)$
\For{$district_{i}$ in $districts$}
    \For{$district_{j}$ in $districts$}
        \State $nodes_{i},nodes_{j}$ = $sample\_nodes(district_{i},district_{j},edges\_btw\_districts_{i,j})$
        \State $disjoint\_union\_graph = connect\_edges(nodes_{i},nodes_{j},disjoint\_union\_graph)$
    \EndFor 
\EndFor \\
\Return $disjoint\_union\_graph$
\EndProcedure
\end{algorithmic}
\end{algorithm}

\begin{algorithm}[H]
\caption{
\texttt{Epidemic spread model} \\
$num\_time\_of\_day$: 4 divisions within a day: morning, noon, evening and night. \\
$set\_time(.)$: Initializes incubation, recovery and mortality time for every node. \\
$rand(a, b)$: Returns a random number between $a$ and $b$. \\
$get\_exp\_inf\_nodes(.)$: Returns exposed and infected nodes eligible for recovery. \\
$get\_exp\_nodes(.)$: Returns exposed nodes eligible for infection. \\
$get\_scp\_nodes(.)$: Returns susceptible nodes eligible for exposure. \\
$get\_inf\_nodes(.)$: Returns infected nodes eligible for death. \\
$get\_lockdown\_flags(.)$: Returns district-wise lockdown flags on the basis of the current network state. \\
$\alpha$ (infection rate) = $ 1 / T_{I}$ \\
$\gamma$  (recovery rate from infection) = $ 1 / T_{RI}$ \\
$\delta$ (recovery rate from exposure) = $ 1 / T_{RE}$
}
\label{alg:EpidemicModel}
\begin{algorithmic}[1]
\Procedure{$Gillespie_{SEIRD}$}{$N$, $I_{init}$, $\beta_{E}$, $\beta_{I}$, $\alpha$, $\delta$, $\gamma$, $\mu$, $T_{I}$, $T_{RI}$, $T_{RE}$, $T_{M}$, $C_{I}$, $C_{E}$, $C_{BS}$, $P_{re}$, $P_{ce}$,, $M_{DD}$, $M_{ZZ}$, $T_{max}$}
\State $num\_time\_of\_day = 4$
\State $N$ =  $set\_time(N, T_{I}, T_{RI}, T_{RE}, T_{M})$
\State $N$, $infected\_nodes$ =  $set\_initial\_infections(N, I_{init})$
\State $exposed\_nodes$ = $get\_exposed\_nodes(N)$
\State $nw\_params$ = $get\_nw\_params(N, exposed\_nodes, infected\_nodes, \beta_{E}, \beta_{I}, \alpha, \delta, \gamma, \mu)$
\State $day\_count$ = 0
\State $nw\_trajectory = [ ]$
\While {$day\_count < T_{max}$ and $nw\_params.tot\_rt > 0$}
    \State $day\_count += 1$
    \State $time\_of\_day = 0$
        \While {$time\_of\_day < num\_time\_of\_day$}
            \State $time\_of\_day += 1$
            \State $ct$ = $get\_current\_time(day\_count, time\_of\_day)$
            \If{not ($day\_count$ == 1 and $time\_of\_day$ == 1)}
                \State $N$ = $mobilize(N, day\_count, time\_of\_day, C_{I}, C_{E}, C_{BS}, P_{re}, P_{ce}, M_{DD}, M_{ZZ}, lockdown\_flags)$
                \State $nw\_params$ = $get\_nw\_params(N, exposed\_nodes, infected\_nodes, \beta_{E}, \beta_{I}, \alpha, \delta, \gamma, \mu)$
            \EndIf
            \State $r$ = $rand(0, 1)$
            \State $P_{R}$ = $\frac{nw\_params.tot\_rec\_rt}{nw\_params.tot\_rt}$
            \State $P_{I}$ = $\frac{nw\_params.tot\_inf\_rt}{nw\_params.tot\_rt}$
            \State $P_{E}$ = $\frac{nw\_params.tot\_exp\_rt}{nw\_params.tot\_rt}$
            \State $P_{D}$ = $\frac{nw\_params.tot\_mor\_rt}{nw\_params.tot\_rt}$
            
            \If{$r < P_{R}$}
                \State $exp\_inf\_nodes$ = $get\_exp\_inf\_nodes(P_{\mu}, F_{sym}, exposed\_nodes, infected\_nodes)$
                \For{$ei$ in $exp\_inf\_nodes$}
                    \If{$ei.ep\_st$ == $I$ and $ct - ei.time >= T_{RI}(ei)$}
                        \State $N.ei.ep\_st$ = $R$
                        \State $N.ei.time$ = $ct$
                    \ElsIf{$ei.ep\_st$ == $E$ and $ct - ei.time >= T_{RE}(ei)$}
                        \State $N.ei.ep\_st$ = $R$
                        \State $N.ei.time$ = $ct$
                    \EndIf
                \EndFor
            \ElsIf{$P_{R} <= r < P_{R} + P_{I}$}
                \State $exp\_nodes$ = $get\_exp\_nodes(F_{sym}, nw\_params.inf\_prob\_mapping)$
                \For{$e$ in $exp\_nodes$}
                    \If{$ct - e.time >= T_{I}(e)$}
                        \State $exposed\_nodes.remove(e)$
                        \State $infected\_nodes.append(e)$
                        \State $N.e.ep\_st$ = $I$
                        \State $N.e.time$ = $ct$
                    \EndIf
                \EndFor
                \algstore{myalg}
\end{algorithmic}
\end{algorithm}

\begin{algorithm}[H]                   
\begin{algorithmic}[1]      
\algrestore{myalg}
            \ElsIf{$P_{R} + P_{I} <= r < P_{R} + P_{I} + P_{E}$}
                \State $scp\_nodes$ = $get\_scp\_nodes(nw\_params.exp\_prob\_mapping)$
                \For{$s$ in $scp\_nodes$}
                    \State $exposed\_nodes.append(s)$
                    \State $N.s.ep\_st$ = $E$
                    \State $N.s.time$ = $ct$
                \EndFor
            \Else
                \State $inf\_nodes$ = $get\_inf\_nodes(P_{\mu}, infected\_nodes)$
                \For{$i$ in $inf\_nodes$}
                    \If{$ct - i.time >= T_{M}(i)$}
                        \State $infected\_nodes.remove(i)$
                        \State $N.i.ep\_st$ = $D$
                        \State $N.i.time$ = $ct$
                    \EndIf
                \EndFor
            \EndIf
            \State $lockdown\_flags$ = $get\_lockdown\_flags(N)$
            \State $nw\_trajectory.append(N)$
            \If{$nw\_params.tot\_rt <= 0$}
                \State break
            \EndIf
        \EndWhile
        \If{$nw\_params.tot\_rt <= 0$}
                \State break
        \EndIf
\EndWhile \\
\Return $nw\_trajectory$
\EndProcedure
\end{algorithmic}
\end{algorithm}

\begin{algorithm}[H]
\caption{\texttt{Mobility model} \\
$isolate\_infected(.)$: Isolates infected nodes with a compliance probability $C_{I}$ and sets $is\_free$ attribute to 0. \\
$isolate\_exposed(.)$: Isolates exposed nodes with a compliance probability $C_{E}$ and sets $is\_free$ attribute to 0. \\
$isolate\_dead(.)$: Isolates dead nodes and sets $is\_free$ attribute to 0. \\
$seal\_borders(.)$: Disconnects districts on the basis of $lockdown\_flags$ with a compliance probability $C_{BS}$. \\
$contact\_tracing\_and\_isolation(.)$: Traverses and tests the neighbourhood of infected agents and isolates them if tested positive. \\
$get\_lckdwn\_flag(.)$: Returns lockdown flag for the current district. \\
$get\_mat(.)$: Returns the appropriate matrix with movement probabilities on the basis of day of week, current time and lockdown. \\
$get\_target(.)$: Returns the target district and zone on the basis of $mvmnt\_mat$ and agent features like employment status etc. It also ensures that at night, agents return to their respective residential districts with probability $C_{H}$ and inter family edges are removed with a probability $P_{fe}$.\\
$update\_edges(.)$: Updates network interactions on the basis of target district and target zone. In case of short range movement(within the same zone of the same district), single edge is removed with $P_{re}$ and another is created with $P_{ce}$. Otherwise, all edges are removed with $P_{re}$ and a single edge is created in the target area with $P_{ce}$. No changes are made if the node does not move. \\
$update\_spatial\_config(.)$: Updates latitude-longitude and zonal coordinates as per the movement.}
\label{alg:MobilityModel}
\begin{algorithmic}[1]
\Procedure{$mobilize$}{$N, day\_count, time\_of\_day, C_{I}, C_{E}, C_{BS}, P_{re}, P_{ce}, M_{DD}, M_{ZZ}, lockdown\_flags$}
\State $ct$ = $get\_current\_time(day\_count, time\_of\_day)$
\State $N$ = $isolate\_infected(N, C_{I})$
\State $N$ = $isolate\_exposed(N, C_{E})$
\State $N$ = $isolate\_dead(N)$
\State $N$ = $seal\_borders(N, C_{BS}, lockdown\_flags)$
\State $N$ = $contact\_tracing\_and\_isolation(N, T_{LB}, N_{RT})$

\For{$node$ in $N.nodes$}
    \If{$node.is\_free$}
        \If{$node.is\_essential\_service\_provider$}
            \State $mvmnt\_mat$ = $get\_mat(node, ct, 0, M_{DD}, M_{ZZ})$
        \Else
            \State $lckdwn\_flg$ = $get\_lckdwn\_flag(lockdown\_flags, node.curr\_dist)$
            \State $mvmnt\_mat$ = $get\_mat(node, ct, lckdwn\_flg, M_{DD}, M_{ZZ})$
        \EndIf
        \State $target\_dist$, $target\_zone$ = $get\_target(node, mvmnt\_mat, ct, C_{H}, P_{fe})$
        \State $N$ = $update\_edges(node, target\_dist, target\_zone, N, P_{re}, P_{ce})$
        \State $N$ = $update\_spatial\_config(node, target\_dist, target\_zone, N)$
    \EndIf
\EndFor \\
\Return $N$
\EndProcedure
\end{algorithmic}
\end{algorithm}

\begin{algorithm}[H]
\caption{\texttt{Get Network Parameters} \\
$num\_neighbors(.)$: Returns number of neighbors of a node in a specific epidemic state.}
\label{alg:GetNwParams}
\begin{algorithmic}
\Procedure{$get\_nw\_params$}{$N, exposed\_nodes, infected\_nodes, \beta_{E}, \beta_{I}, \alpha, \delta,$ $\gamma, \mu$}
\State $nodes\_at\_risk\_of\_exposure$ = $[]$
\State $neighbors\_of\_exposed$ = $N.neighbors(exposed\_nodes)$
\State $neighbors\_of\_infected$ = $N.neighbors(infected\_nodes)$
\State $nodes\_at\_risk\_of\_exposure.append(neighbors\_of\_exposed)$
\State $nodes\_at\_risk\_of\_exposure.append(neighbors\_of\_infected)$
\State $exp\_prob\_mapping$ = $dict()$
\State $tot\_exp\_rt$ = $0$
\For{$node$ in $nodes\_at\_risk\_of\_exposure$}
    \State $er$ = $\beta_{E} \times num\_neighbors(node, E)$ + $\beta_{I} \times num\_neighbors(node, I)$
    \State $exp\_prob\_mapping[node]$ = $er$
    \State $tot\_exp\_rt += er$
\EndFor
\State $inf\_prob\_mapping$ = $dict()$
\State $tot\_inf\_rt$ = $0$
\For{$node$ in $exposed\_nodes$}
    \State $inf\_prob\_mapping[node] = \alpha$
    \State $tot\_inf\_rt += \alpha$
\EndFor
\State $tot\_rec\_rt$ = $\gamma \times len(infected\_nodes) + \delta \times len(exposed\_nodes)$
\State $tot\_mor\_rt$ = $\mu \times len(infected\_nodes)$
\State $tot\_rt$ = $tot\_exp\_rt + tot\_inf\_rt + tot\_rec\_rt + tot\_mor\_rt$ \\
\Return $nw\_params(exp\_prob\_mapping, inf\_prob\_mapping, tot\_exp\_rt,$ $tot\_inf\_rt, tot\_rec\_rt, tot\_mor\_rt, tot\_rt)$
\EndProcedure
\end{algorithmic}
\end{algorithm}

\begin{figure}
    \centering
    \includegraphics[width=\linewidth]{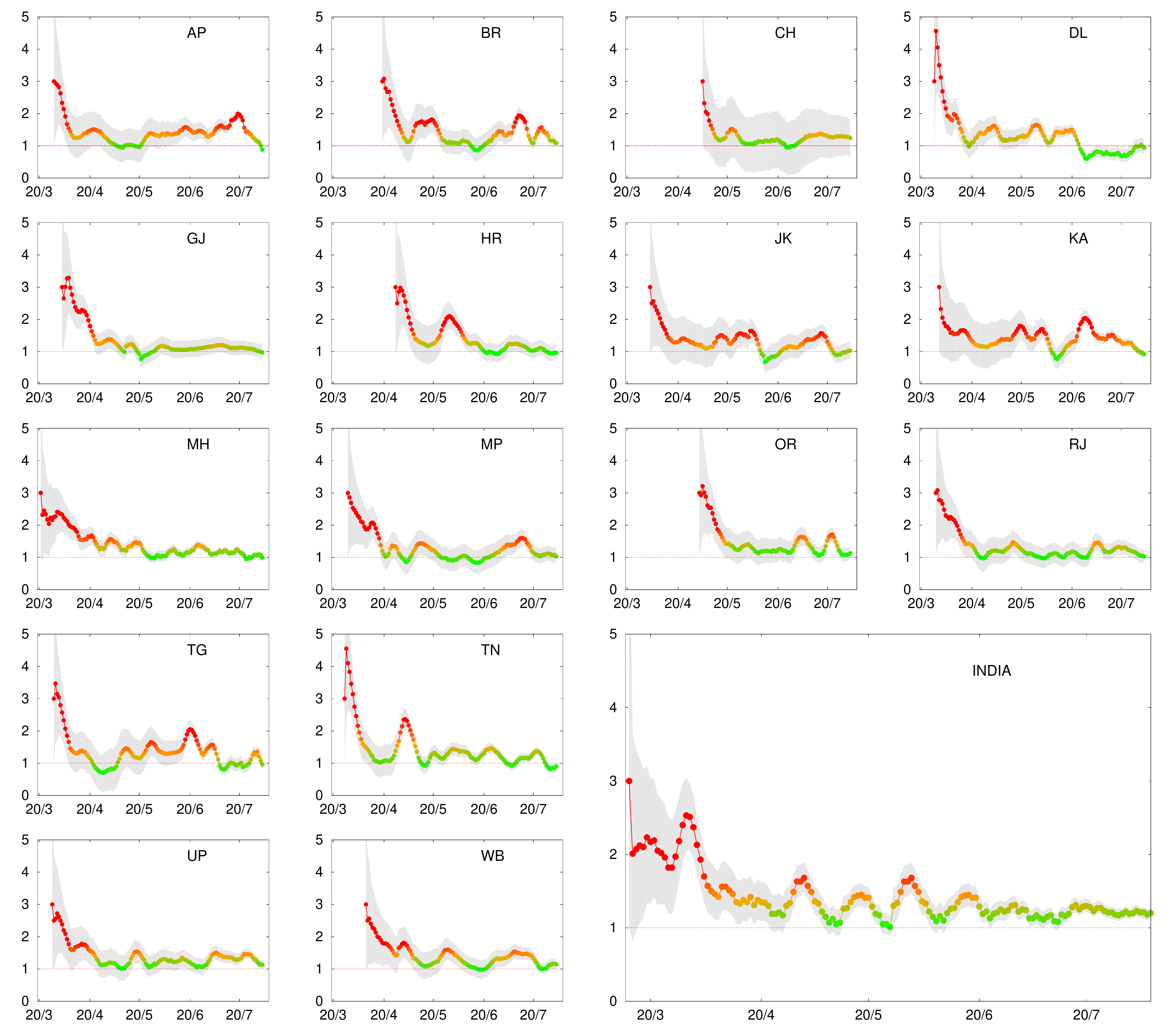}
    \caption{The evolution of the effective reproduction rate $\rho(t)$ with time for some states of India as well as the whole country (shown until 31/07/2020).
    $\rho(t)$ calculated using code from  \texttt{https://github.com/k-sys/covid-19/blob/master/Realtime\%20R0.ipynb} and statewise India data from Covid19india API.}
    \label{fig:rhot_emp}
\end{figure}

\begin{figure}[h]
    \includegraphics[width=\linewidth]{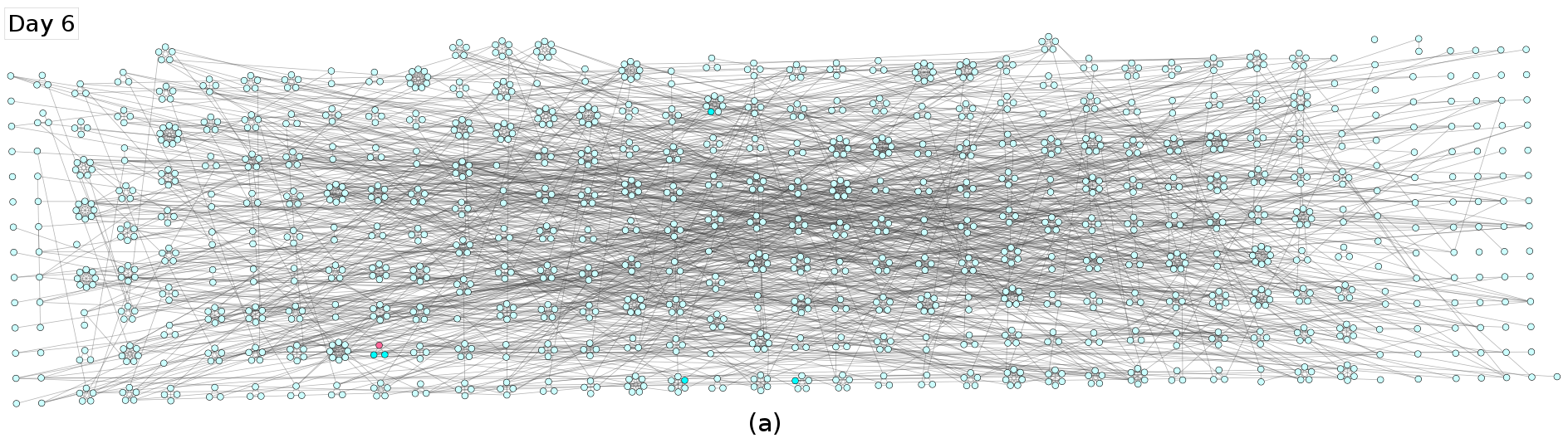}
    \includegraphics[width=\linewidth]{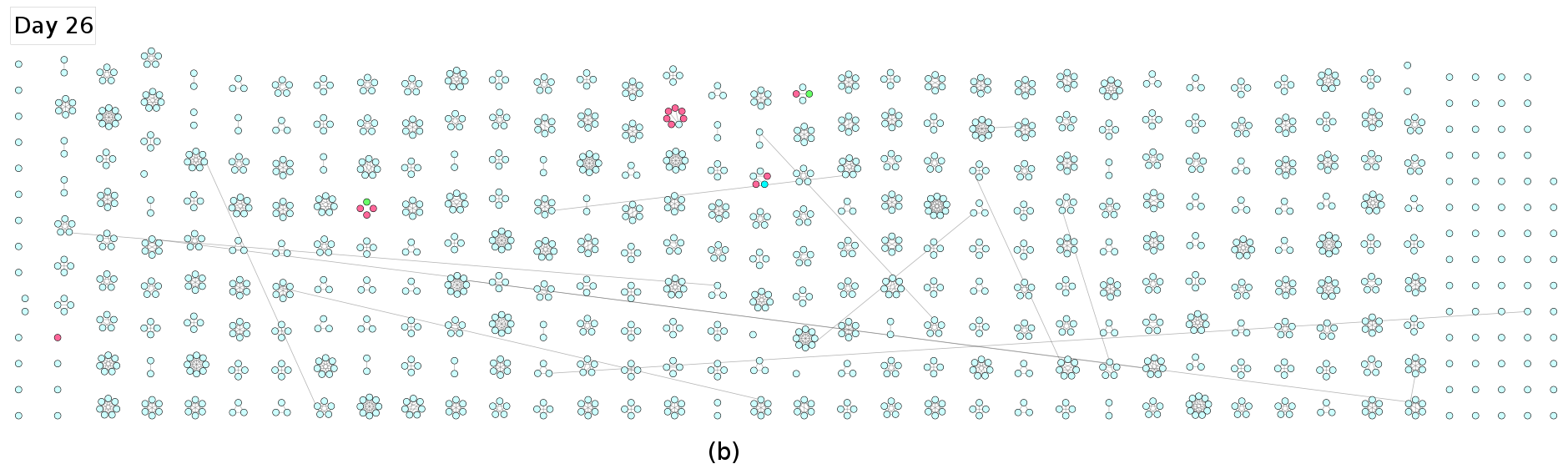}
    \includegraphics[width=\linewidth]{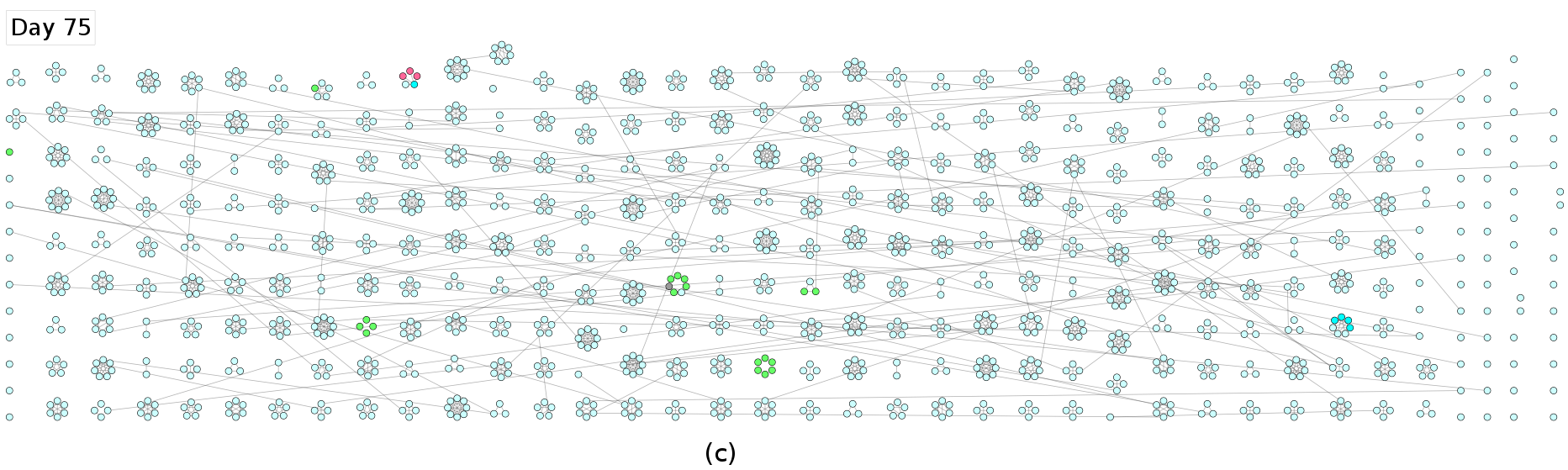}
    \caption{The simulated connected network of families in a single residential zone of \textit{Mumbai suburban} district in Maharashtra during the epidemic on (a) day 6 (b) day 26 and (c) day 75 at night time. Some nodes are absent and new nodes are present due to mobility across zones and districts. The colors depict the epidemic states of the individuals. Susceptible S (cyan), Exposed E (blue), Infected I (red), Recovered R (green), Dead D (grey). This is one of the simulated networks for the data presented in Fig.~\ref{fig:lockdown_sim}.}
     \label{fig:family}
\end{figure}

\end{document}